\documentclass[preprint,12pt]{elsarticle}
\usepackage{amsmath}
\usepackage{amssymb}
\usepackage{eucal}
\usepackage{graphics}
\usepackage{mathtools}
\newcommand{\lgmode}[2]{LG$_{#1}^{#2}$}
\newcommand{\bra}[1]{\langle #1 |}
\newcommand{\ket}[1]{| #1 \rangle}

\newcommand{\ham}{\widehat{H}}

\newcommand{\imag}{\textrm{i}}
\newcommand{\pd}[2]{\frac{\partial #1}{\partial #2}}
\journal{~}
\begin{document}
\begin{frontmatter}
	\title{Electromagnetically induced transparency with Laguerre-Gaussian modes in ultracold rubidium}
	\author{T.~G.~Akin}
	\author{S.~P.~Krzyzewski}
	\author{A.~M.~Marino}
	\author{E.~R.~I.~Abraham}
	\cortext[cor]{Corresponding author.  Tel.: + 1 405 325 2961; Fax: +1 405 325 7557.}
	\ead{abraham@nhn.ou.edu}
	\address{Homer L. Dodge Department of Physics and Astronomy, University of Oklahoma, 440 W. Brooks St. Norman, OK 73019, USA\corref{cor}}
	\begin{abstract}
	We demonstrate electromagnetically induced transparency with the control laser in a Laguerre-Gaussian mode.  	The transmission spectrum is studied in an ultracold gas for the D2 line in both $^{85}$Rb and $^{87}$Rb, where the decoherence due to diffusion of the atomic medium is negligible.
	We compare these results to a similar configuration, but with the control laser in the fundamental laser mode.
	We model the transmission of a probe laser under both configurations, and we find good agreement with the experiment.
	We conclude that the use of Laguerre-Gaussian modes in electromagnetically induced transparency results in narrower resonance linewidths as compared to uniform control laser intensity.
	The narrowing of the linewidth is caused by the spatial distribution of the Laguerre-Gaussian intensity profile.
	\end{abstract}
	\begin{keyword}
		Laguerre-Gaussian mode \sep Electromagnetically Induced Transparency\sep Coherent Control \sep Diffractive optics
	\end{keyword}
\end{frontmatter}
\date{\today}
%
%
\section{Introduction}
%
%
Electromagnetically induced transparency (EIT) is an optical technique used to manipulate quantum states of atoms and photons \cite{fim05}.
A control laser modifies the absorption profile of a probe laser, causing coherent destructive interference of excitation pathways of the atom.
The result is an increased transmission of the probe laser tuned to an atomic resonance where absorption is otherwise expected.
Applications of EIT range from coherent storage of light in the atomic medium for quantum information storage~\cite{ldb01,pfm01}, nonlinear optics~\cite{hfi90}, and lasing without population inversion~\cite{a91}. \\
\indent Initial spectroscopic studies of EIT were performed in an atomic gas at room temperature~\cite{bih91}.
Large laser powers can overcome the Doppler broadening, but cause homogeneous line broadening, though specific Doppler-free techniques produce EIT signals in a room temperature gas with moderate laser powers~\cite{glj95,om09,mhr10,csb11}.
Alternatively, experiments that produce ultracold samples of atomic gases result in Doppler-broadening smaller than the natural linewidth of the atomic transition.
The reduced transverse motion of cold atoms also suppresses the decoherence due to diffusion.
Ultracold gases also offer high densities, typically in the range from $10^{9}-10^{12}$~cm$^{-3}$.
For these reasons, EIT has been extensively studied in this environment~\cite{huc97,yrz01,wzj03,klv09,tsr10}. \\
\indent One consequence of EIT is the slowing of light in an atomic sample.
The destructive interference of excitation pathways in EIT leads to a sub-natural linewidth transmission feature.
There is no theoretical minimum to the linewidth, which is only limited by experimental constraints, such as background magnetic fields, laser linewidth, atomic collisions, and other homogeneous broadening~\cite{fim05}.
Slow light results from enhancement of the slope of the dispersion in the frequency range near the EIT resonance.
Lowering of the intensity of the control field leads to narrowing of the EIT linewidth and results in the decreasing of probe group velocity.
Speeds many orders of magnitude less than $c$ have been achieved in an ultracold gas~\cite{hhd99} and in room temperature gases~\cite{bkr99,ksz99}.
A slowed probe pulse propagating through the medium can be coherently stored in and retrieved from the atoms by adiabatically switching the control laser off and on~\cite{ldb01,pfm01}. \\
\indent Incorporation of a laser propagating in a Laguerre-Gaussian (\lgmode{p}{\ell}) mode to EIT is of considerable interest.
The azimuthal winding phase ($e^{i \ell \phi}$) leads to quantized orbital angular momentum (OAM) of $\ell\hbar$ per photon.
A probe laser carrying OAM generates a manifold of information degrees of freedom, allowing multi-dimensional quantum computing and encryption~\cite{mtt07}.
Storage of \lgmode{p}{\ell} mode probe pulses in gases has been demonstrated in both room temperature gases~\cite{psf07,vch08,wzj08,fly10}, and in ultracold gases~\cite{iky06,mft09,vng13,nvg14}.
The OAM forces the intensity to go to zero at the center, and the additional $p$ radial nodes give rise to ``doughnut'' shaped beams, or even concentric ring intensity patterns and a spatially varying Rabi frequency. \\
\indent A control laser with a large Rabi frequency (as compared to the decoherence rates) increases the signal contrast of the EIT feature, while a small Rabi frequency results in a narrower EIT resonance.
Placing a control laser in an \lgmode{p}{\ell} mode, and aligning the probe laser to the central node causes most of the probed atoms to experience a low control field, resulting in a narrowing of the EIT resonance.
Then, the control laser power can be increased, improving the signal.
Previously, sub-natural linewidths were observed in the EIT transmission spectra using an \lgmode{0}{1} control beam in room temperature gases~\cite{arp10,cn13}.
Remarkably, the \lgmode{0}{1} control laser reduced the EIT linewidth by a factor of 2 as compared to a similar experiment with the control in a Gaussian mode.
It was proposed in ref.~\cite{cn13} that the OAM of the control beam could decrease the decoherences due to transit effects of the room temperature gas.  However, they show that increasing OAM does not increase the narrowing and argue that the reduction in EIT linewidth is due entirely to the spatial dependence of the control Rabi frequency. \\
\indent We measure EIT transmission spectra using an \lgmode{0}{1} laser mode in ultracold atoms prepared in a magneto-optical trap (MOT). We find narrowing of the EIT resonance, and in the ultracold system transit decoherences are negligible.  We measure EIT linewidths with the control beam in the \lgmode{0}{1} mode and the probe in the fundamental Gaussian mode for four different EIT configurations of the D2 line for both $^{85}$Rb and $^{87}$Rb.
We compare this data with EIT spectra where both the probe and the control lasers are in the fundamental Gaussian mode.
A theoretical model is used to analyze each configuration.
We use a density matrix formalism for a six level system composing the two ground-state hyperfine levels and the four excited-state hyperfine levels of the D2 transition in $^{85}$Rb and $^{87}$Rb.
We model the fields as plane waves when the control is in the fundamental mode, and include the spatially varying Rabi frequency when the control is in the \lgmode{0}{1} mode~\cite{ka00,arp10}.
A transmission spectrum is generated from the steady-state solutions of the density matrix equations.
We find good agreement between the model and the experiment.
For both theory and experiment, we observe narrower EIT resonance features with the control laser in the \lgmode{0}{1} mode than with the control laser in the Gaussian mode.
%
%
\section{Experimental Design}
\begin{figure}
	\centering
	\includegraphics[width=90mm]{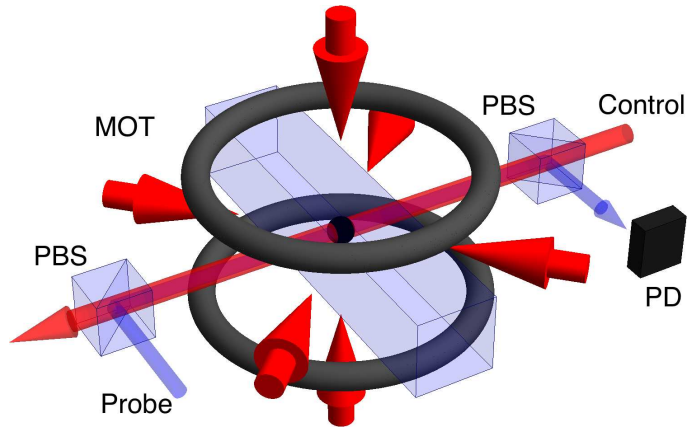}
	\caption{\label{figure:set_up} (Color online.)
	Schematic of the ultracold EIT experimental apparatus.
	The magneto-optical trap (MOT) consists of a pair of anti-Helmholtz coils, six orthogonal and counter-propagating trapping lasers, and a repumping laser (co-propagating with the trapping laser).
	The control laser is indicated by the transparent red beam.
	The probe laser is indicated by the transparent blue beam.
	The control and the probe have orthogonal linear polarizations, and are combined/separated using polarizing beam-splitters (PBS), and are counter-propagating.
	The probe is imaged on a photodiode (PD).}
\end{figure}

%
%
The experimental set-up for our MOT is shown in Fig.~\ref{figure:set_up} and is similar to that found in \cite{kbf14}.
The trapping laser is a low-powered external-cavity diode laser locked $\simeq 15$~MHz to the red of the $\ket{F=3}\rightarrow\ket{F'=4}$ ($\ket{F=2}\rightarrow\ket{F'=3}$) atomic transition in $^{85}$Rb ($^{87}$Rb) as shown in Fig.~\ref{figure:six_levels}.
It is amplified by a tapered amplifier in the master-oscillator power-amplifier configuration and spatially filtered using a polarization maintaining single-mode fiber.
The output of the fiber has a power of 175~mW, and is telescoped to a $1/e^2$ beam diameter of 2.5~cm.
It is further split into  six beams, three of the beams are directed toward the cell along three orthogonal axes, and the other three beams counter-propagate along these axes with opposite circular polarizations.
The repumping laser is locked on resonance with the $\ket{F=2}\rightarrow\ket{F'=3}$ ($\ket{F=1}\rightarrow\ket{F'=2}$) atomic transition in $^{85}$Rb ($^{87}$Rb) as shown in Fig.~\ref{figure:six_levels}.
The laser has a power of 10~mW, is telescoped to a $1/e^2$ diameter of 2.5~cm, and is directed through a polarizing beam-splitter (PBS) to co-propagate with the trapping laser.
The MOT routinely traps $10^8$ atoms at a temperature of $50~\mu$K.
The density distribution of the atoms is Gaussian with a peak density of $(1-5) \times 10^{10}~\textrm{cm}^{-3}$ and $1/e^2$ radius of $\simeq2$~mm.
Three sets of Helmholtz coils eliminate the effects of background magnetic fields. \\
\indent The probe laser is also an external cavity diode laser.
The frequency of the probe is scanned $150-500$~MHz across the resonances indicated in Fig.~\ref{figure:six_levels} by ramping the voltage across a piezoelectric transducer attached to the grating that serves as the output coupler on the external cavity~\cite{wh91}.
The probe laser is shuttered using an acoustic optical modulator (AOM), and is spatially filtered using a polarization maintaining fiber optic cable.
The power after the fiber is approximately $10~\mu\textrm{W}$ and is linearly polarized.
The beam is overlapped with the control laser with a PBS and directed through the center of the MOT (Fig.~\ref{figure:set_up}) with a $1/e^2$ radius of $430~\mu\textrm{m}$.
After passing through the ultracold atomic sample, the probe is separated from the counter-propagating control laser with another PBS and is focused onto a Thorlabs DET200 photodiode (PD) that records the EIT transmission spectrum. \\
\indent The control laser is locked near the transition indicated in each of the four configurations shown in Fig.~\ref{figure:six_levels}.
We use a dichroic atomic vapour laser lock (DAVLL)~\cite{mht07}, resulting in a frequency stability of $\simeq1.5$~MHz.
The laser is shuttered using an AOM, and spatially filtered using a polarization maintaining fiber optic cable.
The output of the fiber is $\simeq 10$~mW.
To select an appropriate Rabi frequency, we attenuate the power using a set of neutral density filters.
The polarization of the control laser is orthogonal to the probe, and the beam is directed counter-propagating to the probe using a PBS (Fig.~\ref{figure:set_up}).
At the MOT, the control laser has a $1/e^2$ radius of 1.1~mm when in the Gaussian mode and $270~\mu$m when in the \lgmode{0}{1} mode.
The \lgmode{0}{1} mode beam waist is smaller than the Gaussian mode waist to increase the peak intensity of the control beam.
This is necessary to counteract the loss of laser power after converting the Gaussian mode to the \lgmode{0}{1} mode.
Since the control laser is a diode laser, the generation of \lgmode{p}{\ell} modes must occur external to the cavity.
We shape the control laser into an \lgmode{0}{1} mode using diffractive optics~\cite{kst02}.
A diffractive optic has microscopic structures etched into the surface using lithography techniques.
Using Huygens' principle, the initial Gaussian mode of the diode laser can be transformed into an \lgmode{p}{\ell} wavefront using two optics.
One optic transforms the intensity profile, and a second optic controls the winding phase.
Diffractive optics can create higher order \lgmode{p}{\ell} modes with high mode purity~\cite{kst02}.\\
\indent The EIT sequence is controlled by a Hewlett-Packard 8175A Digital Signal Generator.
Digital pulses shutter the AOMs on the probe, control, and the repump lasers, and control the current flowing through the MOT coils, which can be turned off in $10~\mu$s.
Initially, the MOT is held in a steady state for 96~ms, and the probe and the control lasers are off.
Next, the magnetic field is switched off and we optically pump the atoms for 1~ms.
The optical pumping procedure depends on the EIT configuration.
For configurations (I) and (II), the repump laser is switched off while the trapping laser optically pumps all the atoms into the lower hyperfine ground-state.
For configurations (III) and (IV), the repump laser stays on to optically pump any atoms from the lower hyperfine ground state to the upper hyperfine ground state.
After optical pumping, the control and probe lasers perform the EIT spectroscopy.
The control laser pulse is 3 ms and precedes the probe laser pulse by 1~ms to prepare the atoms for EIT.
The probe laser is on for 2~ms, during which it scans between $150-500$~MHz, depending on the EIT configuration studied.
A photodiode detects the probe transmission and the signal is read on a TDS-3054 oscilloscope.
The entire procedure takes 0.1~s, and we cycle continuously so the MOT is approximately in a steady state.
%
%
\section{Theoretical Model}
%
%
\begin{figure}
	\centering
	\includegraphics[width=90mm]{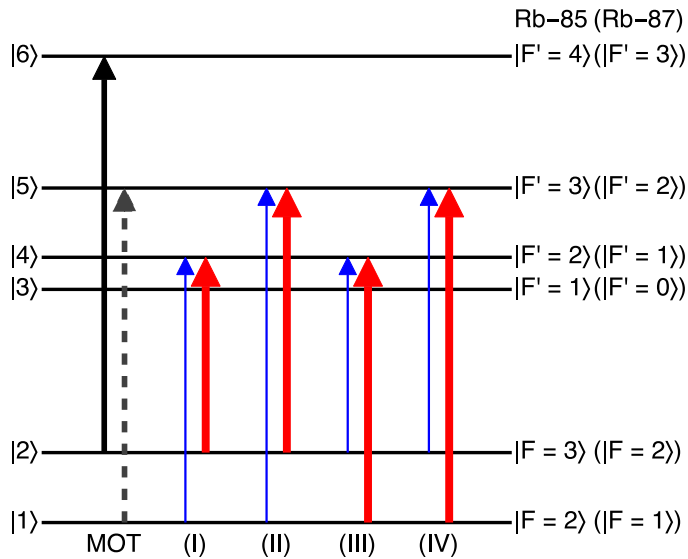}
	\caption{
		\label{figure:six_levels} (Color Online.)
		The six hyperfine states for the D2 transition in $^{85}$Rb ($^{87}$Rb).
		The solid black arrow corresponds to the trapping laser, and the gray dashed arrow corresponds to the repumping laser in our magneto-optical trap (MOT).
		Four EIT configurations on the D2 transition are indicated with Roman numerals.
		The thin blue arrows represent the probe laser, while the thick red arrows represent the control laser.
	}
\end{figure}
We model the experiment using a density matrix formalism~\cite{fim05,arp10,ka00,brg07}.
The density matrix operator is given by $\hat{\rho} = \ket{\psi}\bra{\psi}$, where $\ket{\psi}$ is the atomic state vector.
We consider a six level atom with two levels belonging to the hyperfine levels in the $^2S_{1/2}$ state, and the remaining four levels belonging to the hyperfine levels in the $^2P_{3/2}$ state in rubidium.
Fig.~\ref{figure:six_levels} shows these states and identifies the four EIT configurations we model.\\
\indent The total Hamiltonian for the atom interacting with an electric field is $\widehat{H} = \widehat{H}_0 + \widehat{H}_I$, where the Hamiltonian describing the atomic system is given by:
\begin{equation}
	\label{eq:atom_hamiltonian}
	\widehat{H}_0 = \sum_{n=1}^6 \hbar\omega_n \ket{n}\bra{n},
\end{equation}
where $\hbar \omega_n$ is the energy of the $n$th level.
For each configuration shown in Fig.~\ref{figure:six_levels}, the Hamiltonian describing the interactions with the electric field can be written as:
\begin{table}
	\centering
	\begin{tabular}{r|cccc}
			&	(I)	&	(II)	&	(III)	&	(IV) \\
		\hline
		$i$ &	1	&	1		&	2		&	2	\\
		$k$ &	2	&	2		&	1		&	1	\\
		$l$ &	4	&	5		&	4		&	5
	\end{tabular}
	\caption{\label{table:indices} Appropriate indices for our level scheme.}
\end{table}
\begin{equation}
	\label{eq:interaction_hamiltonian}
\widehat{H}_I = \hbar\Omega_{c,kl}(r)e^{\imag(\omega_{c}t-k_c z +\ell_c\phi)}\ket{k}\bra{l} + \text{h.c.} + \sum_{j=i+2}^{i+4}\left[\hbar\Omega_{p,ij}(r)e^{\imag(\omega_p t +k_p z)}\ket{i}\bra{j}
		+ \text{h.c.}\right],
\end{equation}
where $i$ and $j$ are the level numbers of the states coupled by the probe laser, and $k$ and $l$ correspond to the level numbers of the states coupled by the control laser.
Table~\ref{table:indices} summarizes the level designations that are fixed for each configuration.
The $j$ index is summed over levels for allowed transitions.
In Eqn.~(\ref{eq:interaction_hamiltonian}), $\omega_p$ is the angular frequency and $k_p$ is the wave number of the probe laser.
Likewise, $\omega_c$ is the angular frequency, $k_c$ is the wave number, and $\ell_c$ is the azimuthal charge of the control laser.
The Rabi frequencies for the probe and control lasers are defined as:
\begin{equation}
	\label{eq:rabi}
	\begin{split}
		\Omega_{p,ij}(r) &=-\frac{d_{ij}\mathcal{E}_p(r)}{\hbar}, \\
		\Omega_{c,kl}(r) &=-\frac{d_{kl}\mathcal{E}_c(r)}{\hbar}.
	\end{split}
\end{equation}
The dipole matrix elements, $d_{mn} = \bra{m}\widehat{\epsilon}\cdot\widehat{d}\ket{n}$ of dipole-allowed transitions are taken from refs.~\cite{steckrb85,steckrb87}.
The rest are set to zero.
$\mathcal{E}_p(r)$ and $\mathcal{E}_c(r)$ are the possibly spatially dependent amplitudes of the probe and the control lasers \cite{abs92}:
\begin{equation}
	\label{equation:transverse_field}
	\begin{split}
		\mathcal{E}_p(r) &= \mathcal{E}_{0,p}\text{\hspace{9.1pc}Probe in Gaussian mode} \\
		\mathcal{E}_c(r) &=
			\begin{dcases*}
				\mathcal{E}_{0,c} \left(\frac{\sqrt{2}r}{w_{0,c}}\right)e^{-r^2/w_{0,c}^2} & Control in \lgmode{0}{1} mode\\
				\mathcal{E}_{0,c} & Control in Gaussian mode
			\end{dcases*}
	\end{split}
\end{equation}
where $\mathcal{E}_{0,p}$ and $\mathcal{E}_{0,c}$ are constants, and $w_{0,c}$ is the waist of the control laser.
We define Rabi frequency constants, $\Omega_{0p,ij} = -\frac{d_{ij}\mathcal{E}_{0,p}}{\hbar}$ for the probe laser and $\Omega_{0c,kl} = -\frac{d_{kl}\mathcal{E}_{0,c}}{\hbar}$ for the control laser.
These constants are used to compare different configurations of \lgmode{0}{1} and Gaussian control beams to be consistent with refs.~\cite{arp10,cn13}.
To verify the pane-wave approximation for the Gaussian modes, we evaluated our model with the inclusion of the Gaussian spatial variation.
To the precision of our experiment, we saw no effect on the calculations. \\
\indent We use two unitary transformations on both the density matrix operator and the Hamiltonian.
We transform into the rotating frame using the following unitary transformation:
\begin{equation}
	\widehat{U}_1 = \exp\left[-\imag(\omega_p t + k_p z)\ket{i}\bra{i}-\imag(\omega_c t-k_c z + \ell_c\phi)\ket{k}\bra{k}\right].
\end{equation}
Also, we shift the zero point of the energy to the ground-state hyperfine level indicated by the index, $i$, using:
\begin{equation}
	\widehat{U}_2 =	\exp\left[\imag\omega_p t\ket{i}\bra{i}+\imag\omega_c t\ket{k}\bra{k}\right].
\end{equation}
The transformed Hamiltonian is given by \cite{steck}:
\begin{equation}
	\label{equation:hamiltonian_interaction}
	\ham' = \widehat{U}\ham\widehat{U}^{\dagger}+\imag\hbar\pd{\widehat{U}}{t}\widehat{U}^{\dagger},
\end{equation}
and the density matrix operator is transformed according to $\hat{\rho}' = \widehat{U}\rho\widehat{U}^{\dagger}$.
Applying the rotating wave approximation to the transformed Hamiltonian from Eqn.~(\ref{equation:hamiltonian_interaction}), we arrive at the following full Hamiltonian for our system:
\begin{equation}
	\label{equation:hamiltonian_transform_six_level}
	\begin{split}
		\widehat{H}' = & (-1)^{k}\hbar(\Delta_p - \Delta_c)\ket{k}\bra{k} - \hbar\Delta_p\ket{l}\bra{l}
				- \sum_{j=i+2,j\ne l}^{i+4}\hbar(\Delta_p +\omega_{il}-\omega_{ij}) \\
			&+ \frac{1}{2} \sum_{j=i+2}^{i+4} \hbar\Omega_{p,ij}(\ket{i}\bra{j} + \ket{j}\bra{j})
				+\frac{1}{2} \hbar\Omega_{c,kl}(\ket{k}\bra{l} + \ket{l}\bra{k}),
	\end{split}
\end{equation}
where $\Delta_p = \omega_p - \omega_{il}$ and $\Delta_c = \omega_c - \omega_{kl}$.\\
\indent The time evolution of the density matrix is governed by the Liouville-von Neumann equation \cite{steck}:
\begin{equation}
	\label{equation:master_equation_interaction}
	\imag\hbar\pd{\hat{\rho}'}{t}=\left[\ham',\hat{\rho}'\right]+\widehat{\mathcal{L}} +\widehat{\mathcal{L}}_d,
\end{equation}
where primes indicate a transformed operator, $\widehat{\mathcal{L}}$ is an operator describing the losses in the system, and $\widehat{\mathcal{L}}_d$ describes the dephasing due to the linewidth of the probe and control lasers.
We build the loss operator in a similar fashion to \cite{fim05}:
\begin{equation}
	\label{equation:loss_operator_six}
	\begin{split}
		\widehat{\mathcal{L}} = & \frac{1}{2}\sum_{j=i+2}^{i+3}\Gamma_{ji}\left[2\ket{i}\bra{j}\hat{\rho}' \ket{j}\bra{i}-
			\ket{j}\bra{j}\hat{\rho}' -\hat{\rho}' \ket{j}\bra{j}\right] \\
			& +\frac{1}{2} \sum_{m=4}^5\Gamma_{mk}\left[2\ket{k}\bra{m}\hat{\rho}' \ket{m}\bra{k}
			-\ket{m}\bra{m}\hat{\rho}'-\hat{\rho}' \ket{m}\bra{m}\right],
	\end{split}
\end{equation}
where $\Gamma_{ji}$ is the spontaneous decay rate from state $j$ to $i$.
The decoherence induced by the finite linewidth of the laser is given by \cite{glj95,tpm11}:
\begin{equation}
	\label{equation:decoherence}
	\widehat{\mathcal{L}}_d = - \sum_{m=1}^6\sum_{n=1}^6 \gamma_{mn}\ket{m}\bra{m}\hat{\rho}'\ket{n}\bra{n},
\end{equation}
where the dephasing rate, $\gamma_{mn}$, is the sum of the relevant laser linewidths connecting state $\ket{m}$ to $\ket{n}$.
If we let $\gamma_p$ and $\gamma_c$ be the linewidths of the probe and control lasers, the dephasing rates are:
\begin{equation}
	\begin{split}
		 \gamma_{i,i+2} &= \gamma_{i+2,i} = \gamma_{i,i+3} = \gamma_{i+3,i} = \gamma_{i,i+4} = \gamma_{i+4,i} = \gamma_p, \\
		 \gamma_{kl} &= \gamma_{lk} = \gamma_c, \\
		 \gamma_{12} &= \gamma_{21} = \gamma_p + \gamma_c.
	\end{split}
\end{equation}
All other dephasing rates are zero due to electric dipole selection rules.\\
\indent We make two more assumptions in our model.
The first is that the system is in steady state, which gives $\pd{\hat{\rho}'}{t} = 0$.
The second is that the probe laser is significantly weaker than the control laser.
In this approximation, the density matrix elements ($\rho'_{nm}=\bra{n}\hat{\rho}'\ket{m}$) simplify: $\rho'_{ii} \approx 1$, $\rho'_{nn} = 0$ for $n\ne i$, and $\rho'_{n,n+1} = \rho'_{n+1,n} = 0$ for $n = 1\textellipsis 5$.
Applying these assumptions and solving Eqn.~(\ref{equation:master_equation_interaction}) for the off-diagonal density matrix elements associated with the three dipole-allowed transitions for the probe gives:
\begin{equation}
	\label{eq:solutions}
	\begin{split}
		 \rho'_{il}(r)&=\frac{\frac{1}{2}\Omega_{p,il}(r)\left(\Delta_p-\Delta_c-\imag(\gamma_p+\gamma_c)\right)}
			 {\left(\Delta_p-\frac{\imag}{2}(\Gamma_{li}+\Gamma_{lk}+2\gamma_p)\right)\left(\Delta_p-\Delta_c
			-\imag(\gamma_p+\gamma_c)\right)-\frac{1}{4}\Omega^2_{c,kl}(r)},~\text{and} \\
		\rho'_{ij}(r)&=\frac{\frac{1}{2}\Omega_{p,ij}(r)}{\Delta_p+\omega_{il}-\omega_{ij}
			-\frac{\imag}{2}(\Gamma_{ji}+2\gamma_p)},~\text{for $j = i + 2\textellipsis i + 4$ and $j \ne l$.}
	\end{split}
\end{equation}
The density matrix elements, $\rho'_{ij}(r)$, corresponding to $j\ne l$ behave like two-level transitions.
However, the presence of the strong control laser that couples state $\ket{k}$ to state $\ket{l}$ leads to an EIT feature in the density matrix element $\rho'_{il}(r)$.
Mathematically, this comes from the Rabi frequency term, $\Omega_{c,kl}(r)$, that appears in the expression for $\rho'_{il}$ in Eqn.~(\ref{eq:solutions}).
The $r$-dependence of the density matrix results from the spatial variation of the Rabi frequency.
However, for our system, this dependence is only relevant in the cases involving a control in the \lgmode{0}{1} mode. \\
\indent Our experiment detects the transmission of a weak probe passing through a medium described by the six level model presented above.
Using a semi-classical model, the induced electric dipole moment is given by~\cite{steck}:
\begin{equation}
	\label{dipole}
	p(r) = N\sum_{j=i+2}^{i+4} d_{ij}\left(\rho'_{ij}(r)\right)^{*} = \epsilon_0 \chi(r) \mathcal{E}_p(r),
\end{equation}
where $N$ is the number density of the atomic ensemble which we approximate to be constant, and $\chi(r)$ is the susceptibility.
From Eqn.~(\ref{dipole}), the susceptibility is given by:
\begin{equation}
	\label{equation:susceptibility}
	\chi(r) = \frac{N}{\epsilon_0\mathcal{E_p}(r)}\sum_{j=i+2}^{i+4} d_{ij} \left(\rho'_{ij}(r)\right)^{*}
			= -\frac{N}{\epsilon_0\hbar}\sum_{j=i+2}^{j+4} \frac{d_{ij}^2\left(\rho'_{ij}(r)\right)^{*}}{\Omega_{p,ij}(r)}.
\end{equation}
The imaginary part of the susceptibility describes the absorption of the laser passing through the medium.
Applying Beer's Law to determine the transmitted fraction of the probe laser:
\begin{equation}
	\label{beer}
		T = \frac{\int_0^{w_{0,p}}I_{0,p}(r)e^{-\text{OD}(r)}r\,dr}{\int_0^{w_{0,p}}I_{0,p}(r)r\,dr}
			= \frac{2}{w_{0,p}^2}\int_0^{w_{0,p}}\exp\left(-\text{Im}\left[\chi(r)\right]k_p z\right)r\,dr,
\end{equation}
\begin{figure}
	\centering
	\includegraphics[width=90mm]{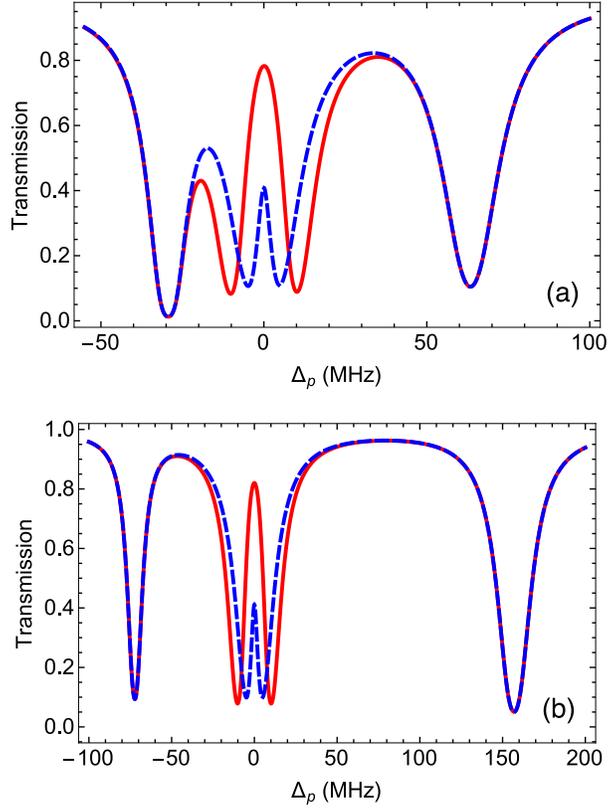}
	\caption{\label{figure:theory}
	(Color online.)  Theoretical curves of the transmission of the probe laser for configuration (I) EIT in (a) $^{85}$Rb and (b) $^{87}$Rb.
	The solid red curve indicates the control laser is in a Gaussian mode, and the dashed blue curve indicates the control laser in an \lgmode{0}{1} mode.
	In both scenarios, the control Rabi frequency is the same ($\Omega_{0c,kl} = 20$~MHz).
	The linewidth of the control laser is assumed to be $\gamma_c = 1$~MHz, the linewidth of the probe laser is set $\gamma_p = 0.1$~MHz, the number density is set to $N = 5\times 10^{10}~\textrm{cm}^{-3}$, and the detuning of the control laser is set to $\Delta_c = 0$.
	Due to the spatial dependence of a \lgmode{0}{1} mode laser, the model predicts that the EIT resonance with the same $\Omega_{0c,kl}$ is narrower for  a control laser in the \lgmode{0}{1} mode than for a control laser in the Gaussian mode.}
\end{figure}
where $\text{OD}(r) = \text{Im}\left[\chi(r)\right]k_p z$ is the optical density, $z$ is the length of the atomic medium along the propagation direction, and the probe laser is assumed to be a plane wave.
We let $z = 4$~mm, which is the $1/e^2$ diameter of the MOT.
We integrate over the region of the EIT interaction, which we take to be a circle with a radius equal to the waist of the probe laser.
We fit Eqn.~(\ref{beer}) to our data using $\Omega_{0c,kl}$, $\gamma_p$, $\gamma_c$, $\Delta_c$, and $N$ as fitting parameters.
The width of the EIT feature is determined from the model by measuring the full-width at half max (FWHM). \\
\indent As a demonstration of the model, theoretical spectra for configuration (I) are shown for $^{85}$Rb in Fig.~\ref{figure:theory}~(a) and for $^{87}$Rb in Fig.~\ref{figure:theory}~(b).
The dashed blue curve is a theoretical spectrum when the control laser is in the \lgmode{0}{1} mode, and the solid red curve is a theoretical spectrum when the control laser is in the Gaussian mode.
For all curves, we assume the control Rabi frequency constant to be $\Omega_{0c,kl} = 20$~MHz, the control laser linewidth to be $\gamma_c = 1$~MHz, the probe laser linewidth to be $\gamma_p = 0.1$~MHz, the number density to be $N = 5\times10^{10}~\textrm{cm}^{-3}$, and the control laser detuning to be $\Delta_c = 0$.
Our model agrees with previous results~\cite{cn13} that for control lasers with equal $\Omega_{0c,kl}$, EIT involving a control laser in the \lgmode{0}{1} mode results in a narrowing of the resonance feature.
%
%
%
\section{Results}
%
%
\begin{figure}
	\centering
	\includegraphics[width=90mm]{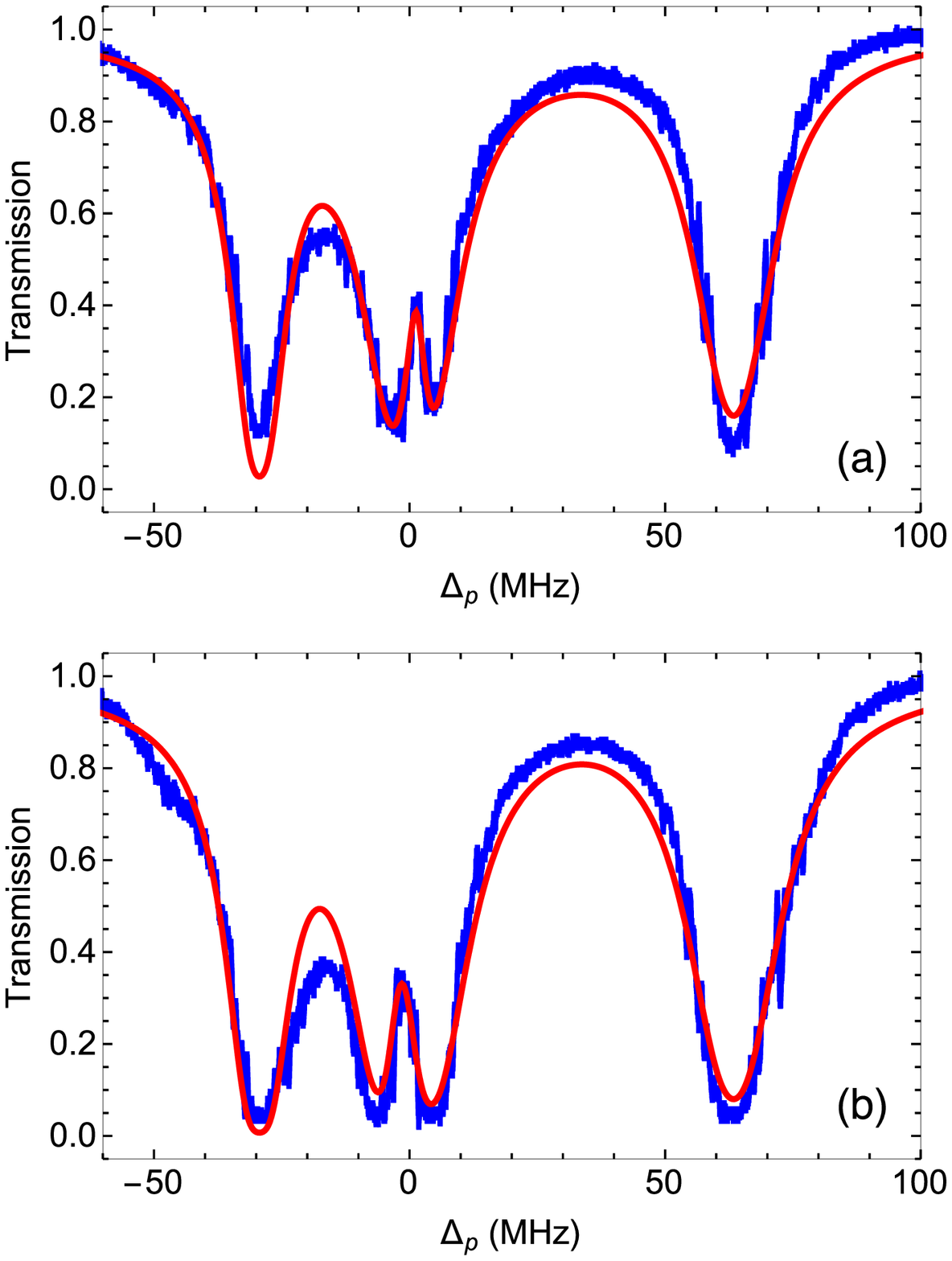}
	\caption{\label{figure:rb85_i} (Color Online.)
	The transmission spectrum for a probe laser scanning over the transitions $\ket{F=2}\rightarrow\ket{F'=1,2,3}$ in $^{85}$Rb.
	The blue curve is the observed signal, and the red curve is the model.
	(a) The control laser is in the \lgmode{0}{1} mode 1.4~MHz above the $\ket{F=3}\rightarrow\ket{F'=2}$ transition with $\Omega_{0c,32} = 16$~MHz.
	(b) The control laser is in the Gaussian mode 1.8~MHz below the $\ket{F=3}\rightarrow\ket{F'=2}$ transition with $\Omega_{0c,32} = 9.5$~MHz.}
\end{figure}
Spectra for all four EIT configurations are measured for $^{87}$Rb, and configurations (I) and (II) are measured for $^{85}$Rb.
Figure~\ref{figure:rb85_i}~(a) shows a measurement of the EIT spectra of $^{85}$Rb when the control is in the \lgmode{0}{1} mode, and Fig.~\ref{figure:rb85_i}~(b) shows the same measurement with the control in the Gaussian mode.
A best fit from the model is also shown in Figs.~\ref{figure:rb85_i}~(a) and (b).
The control laser is locked near the $\ket{F = 3} \rightarrow \ket{F' = 2}$ hyperfine transition, while the probe scans over the $\ket{F = 2} \rightarrow \ket{F' = 1,2,3}$ dipole allowed transitions.
The EIT feature occurs in the $\ket{F = 2} \rightarrow \ket{F' = 2}$ transmission peak, which corresponds to configuration (I) in Fig.~\ref{figure:six_levels}.
We find excellent agreement between the model and data.
We determine the EIT characteristics from the model.
In Fig.~\ref{figure:rb85_i}~(a), the control laser is in the \lgmode{0}{1} mode with $\Omega_{0c,32}=16$~MHz, and the laser frequency is 1.4~MHz above the $\ket{F=3}\rightarrow\ket{F'=2}$ transition.
The FWHM of the EIT feature is $0.67\Gamma$, where $\Gamma = 2 \pi \times 6.07$~MHz.
In Fig.~\ref{figure:rb85_i}~(b), the control laser is in the Gaussian mode with $\Omega_{0c,32} = 9.5$~MHz, and the laser frequency is 1.8~MHz below the $\ket{F=3}\rightarrow\ket{F'=2}$ transition.
The FWHM of the EIT feature is $0.73\Gamma$.
The different values for the control laser frequency detuning are due to a drift in the lock of the control laser. \\
\begin{figure}
	\centering
	\includegraphics[width=90mm]{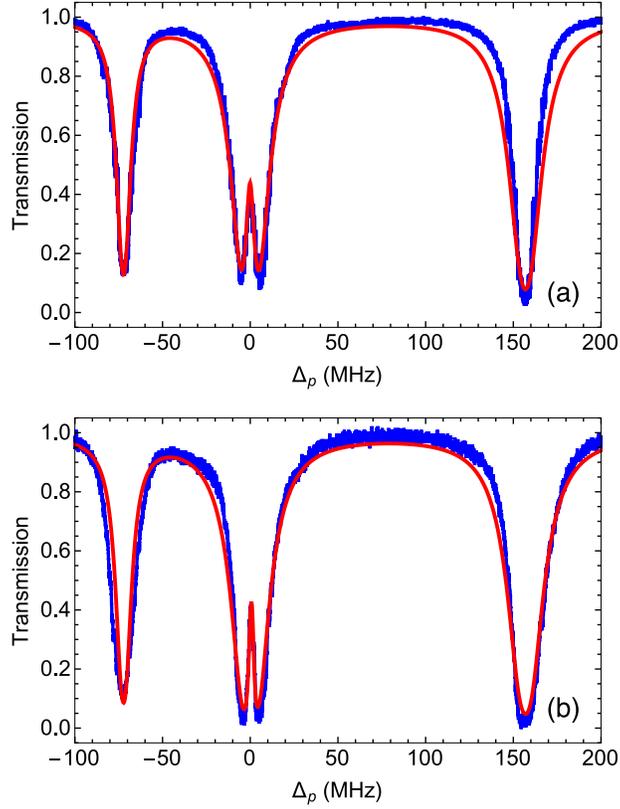}
	\caption{\label{figure:rb87_i} (Color Online.)
	The transmission spectrum for a probe laser scanning over the transitions $\ket{F=1}\rightarrow\ket{F'=0,1,2}$ in $^{87}$Rb.
	The blue curve is the observed signal, and the red curve is the model.
	(a) The control laser is in the \lgmode{0}{1} mode 0.1~MHz below the $\ket{F=2}\rightarrow\ket{F'=1}$ transition with $\Omega_{0c,21} = 20$~MHz.
	(b) The control laser is in the Gaussian mode 0.7~MHz above the $\ket{F=2}\rightarrow\ket{F'=1}$ transition with $\Omega_{0c,21} = 6.7$~MHz.}
\end{figure}
\indent Figure~\ref{figure:rb87_i}~(a) shows a measurement of the EIT spectra of $^{87}$Rb when the control is in the \lgmode{0}{1} mode, and Fig.~\ref{figure:rb87_i}~(b) shows the same measurement with the control in the Gaussian mode.
A best corresponding fit from the model is also shown in Figs.~\ref{figure:rb87_i}~(a) and (b).
The control laser is locked near the $\ket{F = 2} \rightarrow \ket{F' = 1}$ hyperfine transition, while the probe scans over the $\ket{F = 1} \rightarrow \ket{F' = 0,1,2}$ dipole allowed transitions.
The EIT feature occurs in the $\ket{F = 1} \rightarrow \ket{F' = 1}$ transmission peak, which corresponds to configuration (I) in Fig.~\ref{figure:six_levels}.
We again determine the EIT characteristics from the fit.
In Fig.~\ref{figure:rb87_i}~(a), the control laser is in the \lgmode{0}{1} mode with $\Omega_{0c,21} = 20$~MHz, and the laser frequency is 0.1~MHz below the $\ket{F=2}\rightarrow\ket{F'=1}$ transition.
The FWHM of the EIT feature is $0.8\Gamma$.
In Fig.~\ref{figure:rb87_i}~(b), the control laser is in the Gaussian mode with $\Omega_{0c,21} = 9.5$~MHz, and the laser frequency is 0.7~MHz above the $\ket{F=2}\rightarrow\ket{F'=1}$ transition.
The FWHM of the EIT feature is $0.46\Gamma$.
Again, there is excellent agreement between experiment and theory. \\
\begin{table}
	\centering
	\begin{tabular}{r|cccc}
		\hline
		\hline
			Mode			&	Config. 	& $\Omega_{0c,kl}$ (MHz)	& FWHM			& 	 \begin{tabular}{@{}c@{}} Predicted FWHM \\(for opposite mode)\end{tabular}\\
			\hline
			\lgmode{0}{1}	&	(I)			& 16						& $0.67\Gamma$	&	$1.46\Gamma$\\
							&	(II)		& 19						& $0.78\Gamma$	&	$1.87\Gamma$ \\
			\hline
			Gaussian		&	(I)			& 9.5						& $0.73\Gamma$	&	$0.37\Gamma$\\
							&	(II)		& 11						& $0.85\Gamma$	&	$0.41\Gamma$\\
		\hline
		\hline
	\end{tabular}
	\caption{\label{table:rb85} The mode, configuration, $\Omega_{0c,kl}$, and linewidth for EIT experiments in $^{85}$Rb.
	The final column gives the theoretically predicted FWHM of the EIT signal if the control laser is in the opposite mode given in the first column, but same $\Omega_{0c,kl}$.
	Each configuration is described in Fig.~\ref{figure:six_levels}.}
\end{table}
\indent We make similar measurements on the other configurations in Fig.~\ref{figure:six_levels}.
The results for $^{85}$Rb are found in table \ref{table:rb85}, and the results for $^{87}$Rb are in table \ref{table:rb87}.
For $^{85}$Rb, the observed EIT linewidths are the same, but $\Omega_{0c,kl}$ for the control laser for the \lgmode{0}{1} system is 70\% larger.
The larger intensity is necessary to achieve sufficient signal.
Increasing the intensity of the Gaussian control laser for direct comparison would lead to an additional feature in the transmission profile due to effects coming from the degenerate Zeeman sublevels not included in our model~\cite{cly00}.
For $^{87}$Rb, the EIT linewidths we observe are a factor of two larger than those with a Gaussian control beam, but the corresponding values of $\Omega_{0c,kl}$ are 2.5 to 5 times larger.
We also used our model to investigate the predicted linewidths of the EIT for the opposite control mode with same $\Omega_{0c,kl}$ (shown in the final column of Tables~\ref{table:rb85} and \ref{table:rb87}).
The measured linewidths of a configuration with a given $\Omega_{0c,kl}$ in the \lgmode{0}{1} mode are narrower than the predicted linewidths of the identical configuration with the control in the Gaussian mode. Similarly, the predicted linewidths of a configuration with a given $\Omega_{0c,kl}$ of the control in the \lgmode{0}{1} mode are narrower than the measured linewidths of the identical configuration with the control in the Gaussian mode. \\
\indent Given these results and the quality of the fit of the model (Figs.~\ref{figure:rb85_i} and \ref{figure:rb87_i}), both for \lgmode{0}{1} and Gaussian control lasers, we conclude that the linewidth is narrower with the \lgmode{0}{1} control beam for equal $\Omega_{0c,kl}$ in an ultracold gas.
Because the atoms travel $\sim 10~\mu$m during the experiment, the effect is independent of the transit time of the atoms.
This is consistent with what has previously been observed in gases at room temperature where it was shown to be independent of the OAM of the control beam~\cite{arp10,cn13}.
\begin{table}
	\centering
	\begin{tabular}{r|cccc}
		\hline
		\hline
			Mode			&	Config. 	& $\Omega_{0c,kl}$ (MHz)	& FWHM			& 	 \begin{tabular}{@{}c@{}} Predicted FWHM \\(for opposite mode)\end{tabular}\\
			\hline
			\lgmode{0}{1}	&	(I)			& 20						& $0.8\Gamma$	&	$1.94\Gamma$\\
							&	(II)		& 25						& $0.95\Gamma$	&	$2.64\Gamma$\\
							&	(III)		& 33						& $1.5\Gamma$	&	$4.26\Gamma$\\
							&	(IV)		& 33						& $1.2\Gamma$	&	$3.52\Gamma$\\
			\hline
			Gaussian		&	(I)			& 7.3						& $0.46\Gamma$	&	$0.25\Gamma$\\
							&	(II)		& 5.0						& $0.44\Gamma$	&	$0.22\Gamma$\\
							&	(III)		& 6.9						& $0.61\Gamma$	&	$0.29\Gamma$\\
							&	(IV)		& 8.2						& $0.75\Gamma$	&	$0.36\Gamma$\\
		\hline
		\hline
	\end{tabular}
	\caption{\label{table:rb87} The mode, configuration, $\Omega_{0c,kl}$, and linewidth for EIT experiments in $^{87}$Rb.
	The final column gives the theoretically predicted FWHM of the EIT signal if the control laser is in the opposite mode given in the first column, but same $\Omega_{0c,kl}$.
	Each configuration is described in Fig.~\ref{figure:six_levels}.}
\end{table}
%
%
\section{Conclusion}
%
%
We observe EIT transmission spectra using a control beam in an \lgmode{0}{1} laser mode in ultracold atoms prepared in a magneto-optical trap.
We measure EIT linewidths with the control beam in the \lgmode{0}{1} mode, and compare to the spectra generated when the control beam is in the Gaussian mode.
In both mode types, we observe sub-natural linewidths with similar signal contrast.
Our theoretical model for this system shows good agreement with the experiment.
We conclude that EIT in an ultracold gas results in a narrower resonance feature for the same value of $\Omega_{0c,kl}$ when a control beam is in an \lgmode{0}{1} mode as compared to the control beam in a Gaussian mode.
Decoherences due to transit effects are negligible.
Therefore, we conclude that the narrowing of the EIT resonance feature is due to the spatial variation of the Rabi frequency of a control laser in the \lgmode{0}{1} mode. \\
\indent This project is funded by The Research Cooperation.
We would like to acknowledge the constructive conversations with Michael A.~Morrison regarding the development of the theoretical model.
%
%
%
%
%
%
\bibliographystyle{elsarticle-num}
\bibliography{bibliography}
\end{document}